\documentclass[10pt,twocolumn,aps,prb,superscriptaddress]{revtex4-1}

\usepackage[]{natbib}
\usepackage{graphicx}
\usepackage{amssymb}
\usepackage{amsmath}
\usepackage[dvips]{color}

\renewcommand{\vec}[1]{\mathbf{#1}}
\renewcommand{\Re}{\mathop{\mathrm{Re}}}
\renewcommand{\Im}{\mathop{\mathrm{Im}}}
\newcommand{\sign}{\mathop{\mathrm{sign}}}
\newcommand{\ep}{E}
\newcommand{\vep}{\varepsilon}

\setcitestyle{super}

\begin{document}

\title{3D massless Kane fermions observed in a zinc-blende crystal}

\author{M. Orlita}\email{milan.orlita@lncmi.cnrs.fr}
\affiliation{Laboratoire National des Champs Magn\'etiques
Intenses, CNRS-UJF-UPS-INSA, Grenoble, France}
\affiliation{Charles University, Faculty of Mathematics and Physics, Ke Karlovu 5, 121 16 Praha 2, Czech Republic}

\author{D. M. Basko}
\affiliation{Universit\'{e} Grenoble 1/CNRS, LPMMC UMR 5493, B.P.
166, 38042 Grenoble, France}

\author{M. S. Zholudev}
\affiliation{Laboratoire Charles Coulomb (L2C), UMR CNRS 5221,
GIS-TERALAB, Universit\'{e} Montpellier II, 34095 Montpellier,
France} \affiliation{Institute for Physics of Microstructures,
RAS, Nizhny Novgorod, Russia}

\author{F. Teppe}
\affiliation{Laboratoire Charles Coulomb (L2C), UMR CNRS 5221,
GIS-TERALAB, Universit\'{e} Montpellier II, 34095 Montpellier,
France}

\author{W. Knap}
\affiliation{Laboratoire Charles Coulomb (L2C), UMR CNRS 5221,
GIS-TERALAB, Universit\'{e} Montpellier II, 34095 Montpellier,
France}

\author{V. I. Gavrilenko}
\affiliation{Institute for Physics of Microstructures, RAS, Nizhny
Novgorod, Russia}

\author{N. N. Mikhailov}
\affiliation{A.V. Rzhanov Institute of Semiconductor Physics,
Siberian Branch, Russian Academy of Sciences, Novosibirsk 630090,
Russia}

\author{S. A. Dvoretskii}
\affiliation{A.V. Rzhanov Institute of Semiconductor Physics,
Siberian Branch, Russian Academy of Sciences, Novosibirsk 630090,
Russia}

\author{P. Neugebauer}
\affiliation{Institut f\"{u}r Physikalische Chemie,
Universit\"{a}t Stuttgart, Pfaffenwaldring 55, 70569 Stuttgart,
Germany}

\author{C. Faugeras}
\affiliation{Laboratoire National des Champs Magn\'etiques
Intenses, CNRS-UJF-UPS-INSA, Grenoble, France}

\author{A.-L. Barra}
\affiliation{Laboratoire National des Champs Magn\'etiques
Intenses, CNRS-UJF-UPS-INSA, Grenoble, France}

\author{G. Martinez}
\affiliation{Laboratoire National des Champs Magn\'etiques
Intenses, CNRS-UJF-UPS-INSA, Grenoble, France}

\author{M. Potemski}
\affiliation{Laboratoire National des Champs Magn\'etiques
Intenses, CNRS-UJF-UPS-INSA, Grenoble, France}

\maketitle

\onecolumngrid


{\bf Solid state physics and quantum electrodynamics with its
ultra-relativistic (massless) particles meet, to their mutual
benefit, in the electronic properties of one-dimensional
carbon nanotubes as well as two-dimensional
graphene or surfaces of topological insulators. However, clear experimental evidence for electronic states with conical
dispersion relations in all three dimensions, conceivable in
certain bulk materials, is still missing. In the present work, we
fabricate and study a zinc-blend crystal, HgCdTe, at the
point of the semiconductor-to-semimetal topological transition.
Three-dimensional massless electrons with a velocity of about
$10^6\:\mathbf{m/s}$ are observed in this material, as testified
by: (i)~ the dynamical conductivity which increases linearly with
the photon frequency, (ii)~in a magnetic field~$B$, by a
$\sqrt{B}$ dependence of dipole-active inter-Landau-level
resonances and (iii) the spin splitting of Landau levels, which
follows a $\sqrt{B}$ dependence, typical of ultra-relativistic
particles but not really seen in any other electronic system so
far.}\newline\vspace{1mm}

\twocolumngrid

The physics of ``Dirac cones'', which largely dominates the research on
electronic properties of 1D and 2D allotropes of $sp^2$-bonded
carbon~\cite{CharlierRMP07,NovoselovNature05,ZhangNature05} as
well as topological insulators\cite{KonigScience07,HasanRMP10},
is now anticipated to be also explored in 3D solids.
Indeed, there have recently been a number of theoretical
predictions of a class of fairly novel materials with
conical 3D electronic bands, such as Weyl semimetals (with even number
of momentum points where two conical
bands touch) and Dirac semimetals (with one or more momentum points, in which four
conical bands meet). Those compounds, such as the metastable $\beta$-cristobalite
BiO$_2$ (Ref. \onlinecite{YoungPRL12}), pyrochlore iridates such as Y$_2$Ir$_2$O$_7$
(Ref.~\onlinecite{WanPRB11,YangPRB11}),
$A_3$Bi where $A$ is an alkali metal Na, K, or Rb (Ref.~\onlinecite{WangPRB12}),
distorted spinels (Ref.~\onlinecite{SteinbergCM13}),
as well as TlBi(S$_{1-x}$Te$_x$)$_2$ and
TlBi(S$_{1-x}$Se$_x$)$_2$ (Ref. \onlinecite{SinghPRB12}), will be
possibly probed experimentally in the future, to clarify
their bulk electronic structure. The surface states of the
latter compound have already been probed experimentally\cite{XuScience11,SatoNaturePhys11}.

\begin{figure}[b!]
      \includegraphics[width=7.5cm]{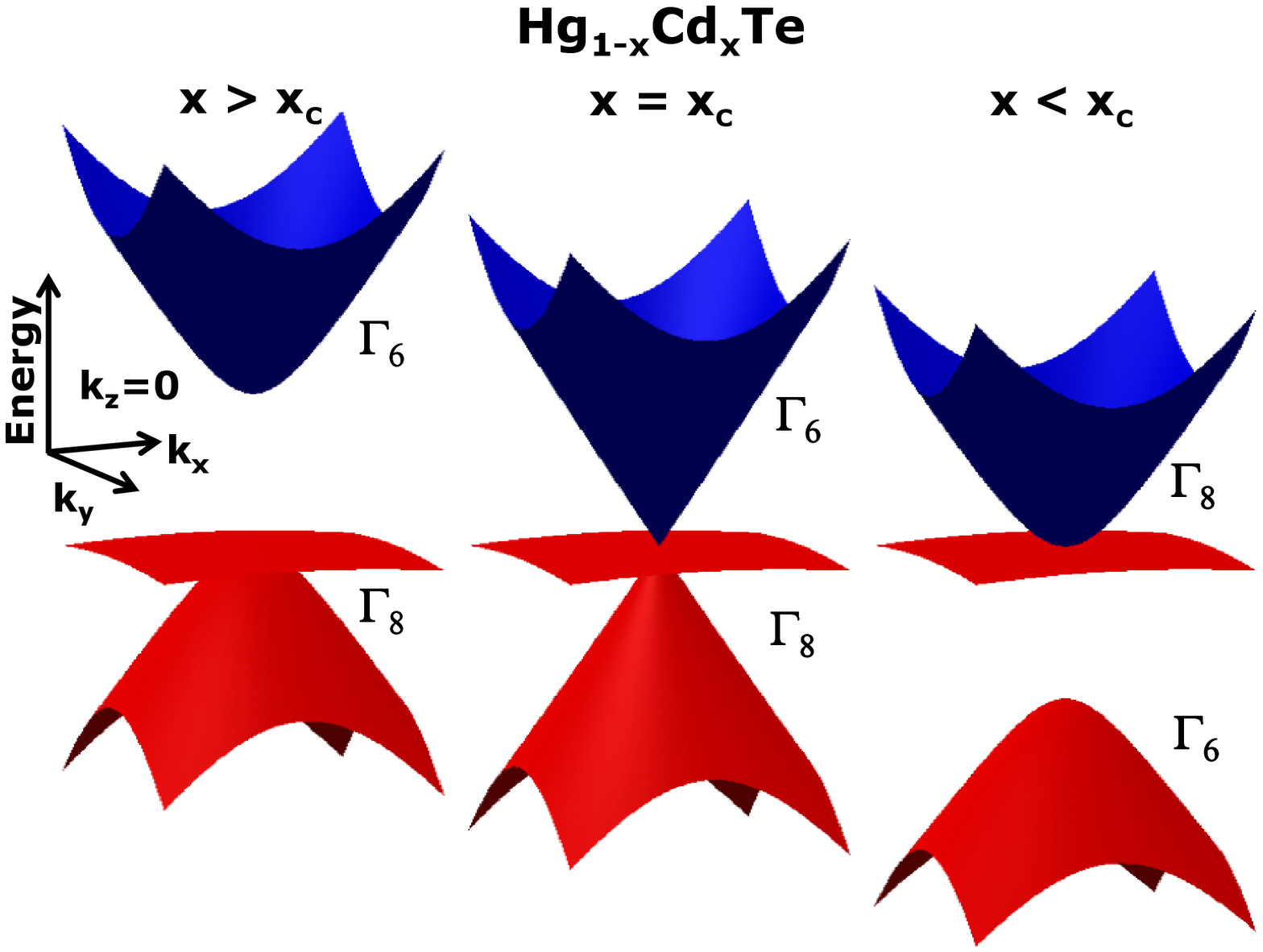}
      \caption{\label{Fig1}
A schematic view of the electronic dispersion of MCT at $k_z=0$
for three different cadmium concentrations~$x$. A standard gapped
semiconductor for $x>x_c$ becomes a semimetal at $x<x_c$. At the point of the topological
transition, $x=x_c$, the conduction band and the light-hole
valence band have a 3D conical dispersion, which is crossed at the
vertex by an almost flat heavy-hole band. In all parts, blue color
corresponds to the conduction band, the valence bands are depicted
in red.}
\end{figure}

On the other hand, the presence of 3D conical dispersion
relations of electronic states was suggested a long time
ago~\cite{ZawadzkiAinP74} to be possible in more conventional, zinc-blende
compounds known as narrow-gap semiconductors. On experimental ground, this particular
shape of dispersion relations has not been seen as the main focus of studies so far.
Nevertheless, a number of investigations, particularly intense in
the sixties and seventies and including magneto-optical studies,\cite{HarmanPRL61,McCombeSSC70,
GrovesSSC71,GuldnerPPSA77,WeilerSaS81} have shown
that the effective mass of the carriers and the energy band gap in Hg$_{1-x}$Cd$_x$Te (MCT)
can be made very small. The band gap was estimated down to tens of meV and the effective mass
to $10^{-2}$ in samples with cadmium concentration close to $x=0.17$. This could be a
sign of linear dispersion relations, though previous experiments studies might have suffered the insufficient
sample quality (inhomogeneous chemical composition and high, unintentional doping) to be more conclusive.

If the cadmium content is sufficiently high, $x>x_c\approx{0}.17$,
the MCT compounds are conventional (narrow gap) semiconductors
with the standard sequence of different symmetry bands: the
$s$-type $\Gamma_6$ band is fixed above the $p$-type $\Gamma_8$
bands, as schematically shown in Fig.~\ref{Fig1}.
Instead, if $x<x_c$, the band order is inverted: the
$\Gamma_6$ band lies below the $\Gamma_8$ bands. As the two
$\Gamma_8$ bands always touch each other at the $\Gamma$ point
of the first Brillouin zone, and only the lower band is
occupied in the intrinsic case, for $x<x_c$ the band structure
is gapless, and MCT becomes a semimetal. The two
distinct phases are not topologically equivalent, as characterized
by a $Z_2$ topological invariant~\cite{BernevigScience06}. At the
point of the topological transition, when the cadmium
concentration reaches its critical value, $x=x_c$, the bandgap
shrinks to zero~\cite{WeilerSaS81} and the electronic dispersion
relation presents some very peculiar properties.

\vspace{1mm}

These conical bands have several spectacular properties similar
to those in Dirac and Weyl semimetals (such as Klein tunnelling
and suppressed backscattering, as discussed below).
Nevertheless, a crucial difference must be stressed. Weyl
semimetals are topologically protected, i.e., the Weyl
points are stable with respect to small perturbations. Dirac
semimetals are not topologically protected, but can be protected
by the crystal symmetry (i.e., the Dirac points are stable with
respect to perturbations which preserve this symmetry). The conical
dispersion in the gapless MCT is not protected by symmetry or
topology; rather, it is achieved by fine tuning of a system parameter
(cadmium concentration). The protected bands might be robust and
then advantageously unaffected by small changes of external parameters.
On the other hand, the band structure of MCT can be suitably
engineered in benefit to design and fabricate the ``gapped-at-will'' compounds
and their interfaces with massless systems.

\vspace{1.5mm}

The basic theoretical approach, needed to understand
these properties, is based on the standard Kane model\cite{KaneJPCS57},
whose validity for MCT has been confirmed by a number of previous studies\cite{HarmanPRL61,McCombeSSC70,
GrovesSSC71,GuldnerPPSA77}
and which usually implies more than 10 free parameters.\cite{WeilerSaS81}
Here, we retain only the terms linear in the wave vector, and neglect the
split-off $\Gamma_7$~band [the magnitude of the splitting, $\Delta\approx{1}\:\mbox{eV}$
(Refs. \onlinecite{WeilerSaS81} and \onlinecite{NovikPRB05})
is assumed to be sufficiently large]. For the remaining six bands,
the Hamiltonian can be written as (see Supplementary Information):
\begin{widetext}\begin{equation}\label{Hk=}
H(\vec{k})=\left(\begin{array}{cccccc}
0 & vk_+\sqrt{3}/2 & -vk_-/2 & 0 & 0 & -vk_z\\
vk_-\sqrt{3}/2 & 0 & 0 & 0 & 0 & 0 \\
-vk_+/2 & 0 & 0 & -vk_z & 0 & 0 \\
0 & 0 & -vk_z & 0 & -vk_-\sqrt{3}/2 & vk_+/2 \\
0 & 0 & 0 & -vk_+\sqrt{3}/2 & 0 & 0  \\
-vk_z & 0 & 0 & vk_-/2 & 0 & 0 \\
\end{array}\right)\equiv{v}\vec{k}\cdot\vec{J},
\end{equation}\end{widetext}
where $k_\pm=k_x\pm{i}k_y$, and the
velocity~$v=\sqrt{E_P/(3m_0)}\approx 10^6\:\mbox{m/s}$ is
expressed in terms of the free electron mass~$m_0$ and the Kane
energy $E_P$ (typically, $E_P\approx 20$~eV for zinc blende
semiconductors, see, e.g., Ref.~\onlinecite{Cardona}). The velocity
is thus the only free parameter, which makes this model extremely simple.

The Hamiltonian (\ref{Hk=}) has three eigenvalues, each
doubly degenerate due to the Kramers theorem (time-reversal
symmetry):
\begin{equation}\label{dispersion=}
\ep_\vec{k}=0,\pm{v}|\vec{k}|
\end{equation}
As usual, the two components of the Kramers doublet can be
labelled by two spin projections $\downarrow,\uparrow$, even
though this degree of freedom has a strong admixture of the
orbital motion due to the spin-orbit coupling. This implies an
anomalously large and nonlinear Zeeman effect.

The eigenvalue $\ep_\vec{k}=0$ corresponds to the heavy-hole band
which, in the approximation of Eq.~(\ref{Hk=}), is dispersionless
(completely flat) or, in other words, characterized by an infinite
effective mass. The inclusion of parabolic terms in the electron
dispersion results in a downward bending of the heavy-hole band,
away from $\vec{k}=0$. This curvature, corresponding to a
heavy-hole mass of about
$m_{hh}\approx{0}.5\,m_0$~\cite{WeilerSaS81}, is not sensitive to
the topological transition at $x=x_c$. The simplified picture of
massless and infinite-mass particles can be used at sufficiently
low energies~$\ep$, such that the ``relativistic'' mass of massless fermions,
$m_c=\ep/v^2\ll{m}_{hh}$. This defines the energy cutoff of
$m_{hh}v^2\approx{3}\:\mbox{eV}$, less stringent than the
spin-orbit splitting $\Delta\approx{1}$~eV. One step beyond the
approximation of Eq.~(\ref{Hk=}) is therefore to use the
eight-band model with a finite~$\Delta$ though still ignoring the
apparent dispersion of the heavy hole band (see Supplementary
Information for details). We follow such an approach when it is
necessary to refine the analysis of the experimental data.

The matrices $\vec{J}=\{J_x,J_y,J_z\}$ which appear in
Eq.~(\ref{Hk=}) do not satisfy the algebra of angular momentum~1,
nor any other closed algebra. Notably, the massless fermions in
MCT are not equivalent to the three-dimensional Dirac electrons in
the ultra-relativistic limit of the quantum electrodynamics (QED).
For example, the Hamiltonian in Eq.~(\ref{Hk=}) has the
characteristic property:
\begin{equation}\label{chirality=}
U_cH(\vec{k})U_c=-H(\vec{k}),\quad
U_c\equiv\mathrm{diag}(1,-1,-1,1,-1,-1).
\end{equation}
Note, however, that there are more $-1$'s than $1$'s in $U_c$,
hence it is not the usual chiral property. As discussed in the
Supplementary Information, the property (\ref{chirality=}) ensures
the existence of a doubly-degenerate flat band. To the best of our
knowledge, the Hamiltonian~(\ref{Hk=}) does not reduce to any
well-known case of massless particles in quantum electrodynamics.
We therefore invoke a new term ``Kane fermions'' to refer to the
electronic states of MCT at the point of the topological
transition (to states in gapless MCT).

Massless Kane fermions share, however, a number of properties with other
ultrarelativistic particles. A prominent example is Klein
tunnelling invoked for 3D Dirac electrons in QED and apparent for
2D Dirac electrons in
graphene~\cite{Katsnelson2006,YoungNaturePhys09}. A perfect
transmission through an arbitrarily high potential barrier at
normal incidence, due to the Klein paradox, should also occur in a
gapless MCT. This can be seen by noting that the eigenstates
corresponding to the same (e.~g., positive) energy but to opposite
wave vectors $\vec{k},-\vec{k}$, also correspond to different
eigenvalues, $\pm{1}$, of the projection of~$\vec{J}$
on~$\vec{k}$. Thus, a potential which does not change $\vec{J}$
(such as the electrostatic potential which acts in the same way on
electrons in all bands and thus is proportional to the unit
matrix), cannot backscatter an electron in the conduction band.

\vspace{2mm}

In order to prove the concept of massless Kane fermions in
experiments, we have used the MBE technique to grow thin layers of
MCT on semi-insulating GaAs substrates (see Supplementary
Information). The optimal structure was used for measurements. It
contains an MCT layer with cadmium concentration close to $x_{\rm
Cd}=0.17$ which extends over a thickness of
$d\approx{3}.2\:\mu\mbox{m}$. The relevant part of this MCT layer
is sufficiently thick to be considered a 3D material and at the
same time thin enough to be suitable for our optical transmission
experiments.


\begin{figure}[t!]
      \includegraphics[width=8cm]{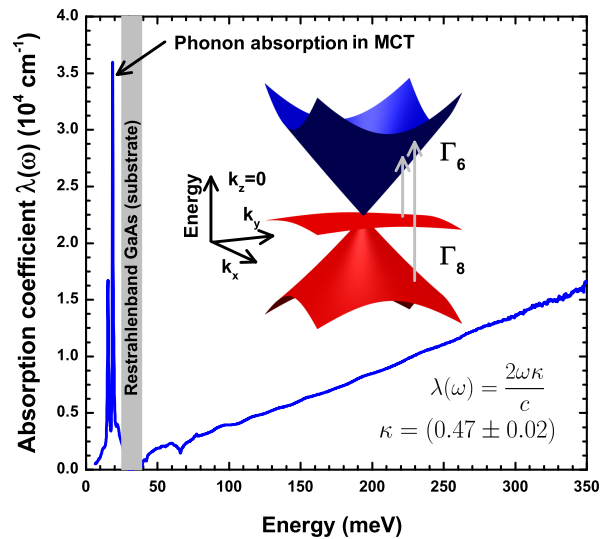}
      \caption{\label{Fig2}
Absorption coefficient of MCT at zero magnetic field,
$\lambda_{B=0}$, measured experimentally, see the text and Methods
for details. While the low-energy response is dominated by
absorption due to phonons, the linear dependence on $\omega$ at
higher photon energies is directly linked to the conical
dispersion of 3D massless fermions. The interband absorption in
gapless MCT, schematically shown in the inset, is dominated by
transitions from the flat (heavy-hole) band.}
\end{figure}

\vspace{1mm}

A striking consequence of conical dispersions on the optical
properties of 3D massless fermions is the absorption coefficient
$\lambda(\omega)$ being proportional to the frequency~$\omega$,
distinctly in contrast to frequency independent absorption of 2D
Dirac electrons as observed in
graphene~\cite{KuzmenkoPRL08,NairScience08}. In simple words,
these characteristic dependences result from the particular forms
of the joint density of states~$\mathcal{D}(\omega)$, which define
the basic absorption profile in solids:
$\lambda(\omega)\propto\mathcal{D}(\omega)/\omega$. A conical
dispersion in 2D yields $\mathcal{D}(\omega)\propto\omega$,
whereas it implies $\mathcal{D}(\omega)\propto\omega^2$, and thus
$\lambda(\omega)\propto\omega$ in case of a 3D system with massless particles.

\begin{figure*}[t]
    \begin{minipage}{0.58\linewidth}
      \includegraphics[width=10.5cm]{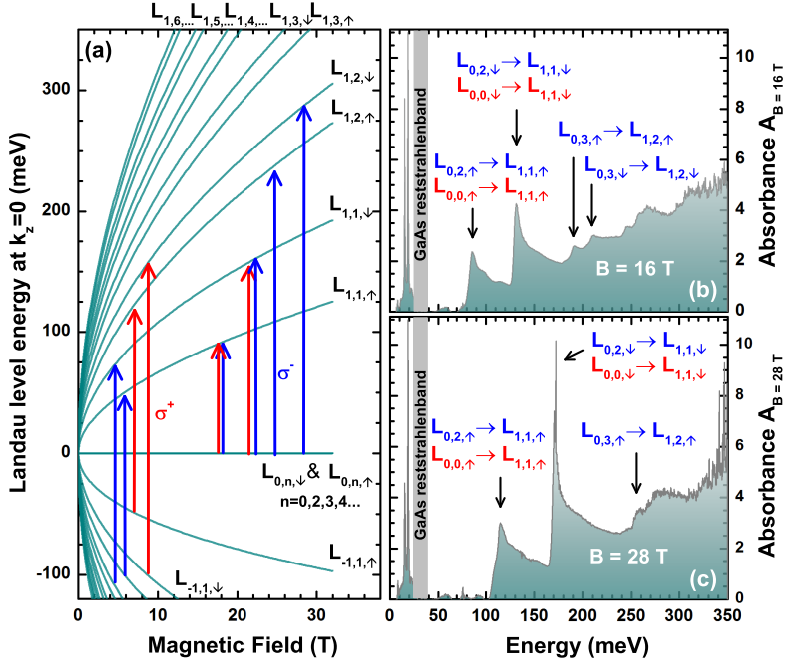}
    \end{minipage}\hfill
    \begin{minipage}{0.38\linewidth}
      \caption{\label{Fig3} Part (a): Landau levels (for $k_z=0$) in gapless MCT, L$_{\zeta,n,\sigma}$,
as a function of the magnetic field, calculated using the eight-band
model, using only $v$ and $\Delta$ parameters. Arrows of
different colors show the optically allowed transitions in undoped gapless MCT
in the two circular polarizations $\sigma^+$ and $\sigma^-$. Parts (b) and (c): Experimentally
measured absorption coefficient $\lambda_B$ (absorbance) as a function of the photon
energy, presented for two values of the magnetic
field, $B=16$ and 28~T, respectively.}
    \end{minipage}
\end{figure*}


To be more quantitative, we use a simple form of the Hamiltonian
given by~Eq.~\eqref{Hk=} and follow the standard
recipe~\cite{Cardona} to analytically derive the dielectric
function $\vep(\omega)$ of the system (see Supplementary
Information). Fixing $\Omega$ for the high-energy cut-off of the
conical dispersion, one finds that if $\omega\ll\Omega$,
then:\cite{Balents}

\begin{equation}\label{vep=}
\vep(\omega)=\vep_\infty +\alpha\frac{13}{12}\,\frac{c}{v}
\left(\frac{2}\pi\ln\frac\Omega{|\omega|}+i\sign\omega\right),
\end{equation}
where $\alpha$ is the fine structure constant ($\alpha$~$\approx$~1/137) and $\vep_\infty$
accounts for the contribution from transitions other than those
described by Eq.~(\ref{Hk=}). The dissipative part of the
dielectric function, $\Im\vep(\omega)$, is dispersionless, while
$\Re\vep(\omega)$ gains a weak (logarithmic) dependence on
$\omega$. Consequently, we derive the dynamical conductivity,
$\sigma(\omega)=i(1-\varepsilon)\varepsilon_0\omega$, the real
part of which is a linear function of~$\omega$. Let us note that
(interband) absorption in gapless MCT is dominated by transitions
from the flat (heavy-hole) band, which is fully occupied in the
intrinsic (undoped) material.

Equation (\ref{vep=}) implies a nearly frequency-independent
extinction coefficient $\kappa=\Im\sqrt{\vep(\omega)}$, and
consequently, the absorption coefficient $\lambda(\omega)$
increasing linearly with $\omega$,
$\lambda(\omega)=2\kappa\omega/c$. The experimentally observed
value $\kappa=0.47\pm{0}.02$ derived directly from the data shown
in Fig.~\ref{Fig2} agrees fairly well with the extinction
coefficient, $\kappa \approx 0.4$, calculated using
Eq.~\ref{vep=}, see Methods for more details.
\vspace{2mm}

In a strong magnetic field, the 3D dispersion is transformed into
a set of Landau levels (LLs), or more precisely, into 1D Landau
bands which disperse with the momentum component along the field
($z$~axis). Inserting the magnetic field into the Hamiltonian in
Eq.~(\ref{Hk=}) via the standard Peierls substitution,
$\hbar\vec{k}\to\hbar\vec{k}-e\vec{A}$, one obtains for gapless MCT
the LL energies [see Eq.~(S28) of the Supplementary Information]:
\begin{equation}\label{LLs=}
\ep_{\zeta,n,\sigma}(k_z)=
\zeta\hbar{v}\sqrt{(2n-1+\sigma/2)l_B^{-2}+k_z^2},
\end{equation}
where $l_B^{-2}={e}B/\hbar$ and the LL index $n=0,1,2\ldots$.
For $n\geq{2}$, the band index is $\zeta=1,0,-1$, while at $n=1$ only
$\zeta=\pm{1}$ are allowed and at $n=0$ only $\zeta=0$ exists.
The states in the flat band remain at zero energy, because the
property~(\ref{chirality=}) remains valid in the presence of a
magnetic field.

The quantum number $\sigma=\pm{1}$ shows how the Kramers
degeneracy, mentioned above, is lifted by the magnetic field.
Thus, $\sigma$~can be viewed as the spin projection on the magnetic
field. The spin splitting is entirely determined by the orbital
parameters $v$, $n$, and $k_z$. Moreover, at $k_z=0$ the spin
splitting of all Landau levels is proportional to $\sqrt{B}$,
which means that the $g$~factor defined in the standard way,
$g_{\zeta,n}=(\ep_{\zeta,n,\uparrow}-\ep_{\zeta,n,\downarrow})/(\mu_B
B)$, diverges at $B\rightarrow0$. This is quite unusual for a
solid state system, and, in particular, does not hold for the
Dirac fermions in graphene. On the other hand, such behavior is
characteristic of ultrarelativistic Dirac electrons in QED,
$E_{\zeta,n,\sigma}(k_z)=
\zeta\hbar{c}\sqrt{(2n+1+\sigma)/l_B^2+k_z^2}$, where
$n=0,1\ldots$ and $\zeta=\pm1$, see, e.g., Ref.~\onlinecite{LL4}.
Note, however, an essential difference: in QED, a level
$(n,\sigma=+1)$ is degenerate with the level $(n+1,\sigma=-1)$.
Such degeneracy is absent for Kane fermions, since it is
$\sigma/2$ that enters Eq.~(\ref{LLs=}). The $\sqrt{B}$ spin
splitting occurs in MCT because the strength of the spin-orbit
coupling becomes effectively infinite when the energy gap
vanishes. Let us now discuss how the $\sqrt{B}$ dependence of
LLs and also of the spin splitting at $k_z=0$, described
by Eq.~(\ref{LLs=}), is verified experimentally.

The magneto-optical response of MCT is determined by
electric-dipole selection rules: $\Delta{n}=n\pm1$ with ``$\pm$''
corresponding to the two circular polarizations, $\Delta{k}_z=0$,
$\Delta\sigma=0$, and no restriction on~$\zeta$. In the undoped
MCT, the incident photon can excite electrons from the filled
valence bands, $\zeta=-1,0$, to the empty conduction band,
$\zeta=1$, as shown schematically in Fig.~\ref{Fig3}(a). Examples
of the measured spectra are shown in Fig.~\ref{Fig3}(b). Since the
dispersion of each Landau band near $k_z=0$ is parabolic, the
joint density of states has sharp inverse-square-root
singularities at energies of the transitions with $k_z=0$, with an
abrupt cutoff on the low-energy side and a shoulder on the
high-energy side, as expected also for Weyl semimetals.\cite{AshbyPRB13}
In the absence of a magnetic field, such a
density of states can be found in 1D Dirac-type systems, in
particular, in carbon nanotubes \cite{CharlierRMP07}. Note that
the singularity of the lowest transition is less sharp than that
of the next one. We attribute this to a small residual electronic
doping, which results in filling of the states with very small
$k_z$ in the lowest Landau level in the conduction band (L$_{1,1,\uparrow}$),
so that the optical transition involving these states is blocked by
the Pauli principle, thereby cutting off the singularity.

\begin{figure}[b]
      \includegraphics[width=8cm]{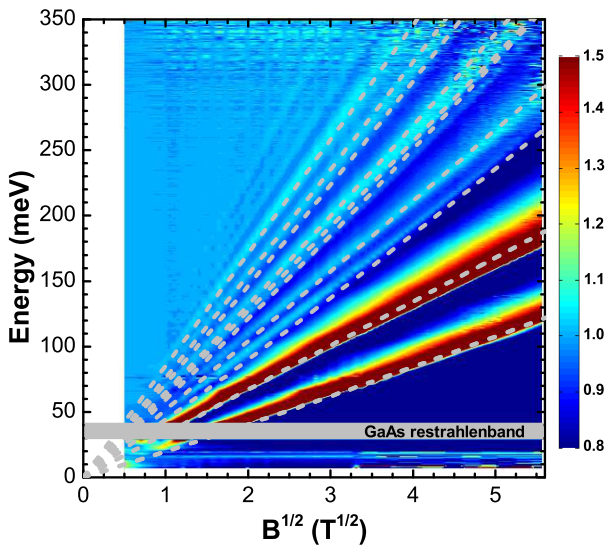}
      \caption{\label{Fig4}
Relative change of absorbance $A_B/A_{B=0}$ plotted
as a false color-map. All the observed resonances clearly follow
$\sqrt{B}$-dependence. The dashed lines are calculated positions
of inter-LL resonances at $k_z=0$ using parameters $v_F=1.06\times
10^{6}$~m/s and $\Delta=1$~eV. The presence of the spin-orbit
split band, expressed by parameter $\Delta$, does not
qualitatively change the LL spectrum, but introduces a weak
electron-hole asymmetry.}
\end{figure}

The key feature of the massless fermions, expressed by
Eq.~(\ref{LLs=}) is the $\sqrt{B}$-dependence of the transition
energies at $k_z=0$. In Fig.~\ref{Fig4}, we plot the infrared
absorbance spectrum (relative to the zero-field absorbance) for
magnetic fields up to 31~T. When plotted as a function of
$\sqrt{B}$, the positions of the $k_z=0$ singularities guide the
eye along straight lines. A close inspection shows that they are
slightly curved at high fields. This weak curvature can be
accounted for by including the spin-orbit split-off band with
$\Delta=1\:\mbox{eV}$. The theoretical curves in Fig.~\ref{Fig4}
were produced using such an 8-band model, considering $v_F$ and $\Delta$ as
only two parameters. The two brightest lines correspond to the transitions from the flat band to the two
spin-split components of the first Landau level in the conduction
band (levels L$_{1,1,\uparrow}$ and L$_{1,1,\downarrow}$, see Fig.~\ref{Fig3}(a)).
This agreement between experiment and theory provides us with
another fingerprint of 3D massless fermions in gapless MCT. Let us note the $\sqrt{B}$-dependence
does not serve as a unique signature of 3D massless particles and it can
be found, for a certain range of magnetic fields,
in the optical response of other bulk materials, e.g., in highly anisotropic
graphite and bismuth\cite{ZhuPRB11,OrlitaPRL08}.

\begin{figure}
      \includegraphics[width=8cm]{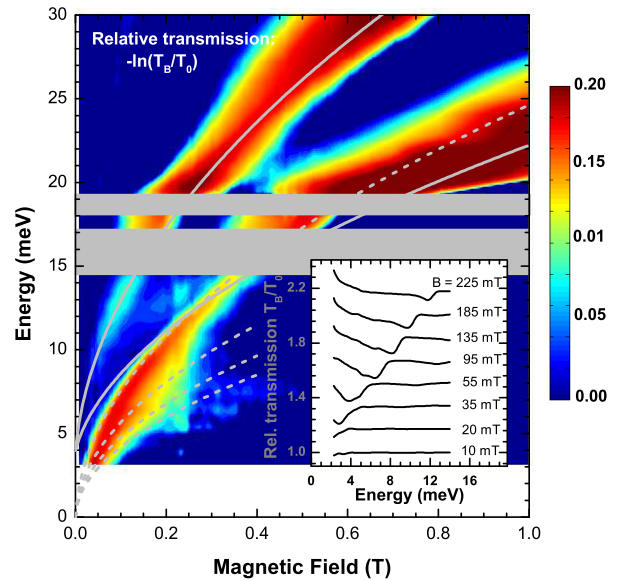}
      \caption{\label{Fig5}
Relative change of transmission, plotted as
$-\ln[T_B/T_0]$, in a form of a color-map at low fields. The inset
shows selected transmission spectra, in which the well-defined CR
response is observed down to 30~mT. The solid and dashed lines
correspond to expected positions of inter-LL transition in a MCT
system with $v=1.06\times10^{6}$~m/s and small energy gap of
$E_g=4$~meV. Solid lines show two lowest lying interband
resonances (from the flat band to L$_{1,1,\uparrow}$ and L$_{1,1,\downarrow}$ levels), the dashed lines
are pure CR-like transitions between pairs of adjacent LLs in the
conduction band ($\zeta=1, n=1,2,3\ldots$). The grey areas
correspond to regions of strong phonon-related absorption in MCT.}
\end{figure}


Having shown that in a wide, $30-300$~meV energy range, the
optical response of our sample is well explained by a model of
gapless and intrinsic (undoped) MCT, we focus now on the low
energy, low magnetic field range (see Fig.~\ref{Fig5}) of our data,
with the aim of estimating the accuracy of such an approach. The spectral
range of interest, below $30$~meV, is not easy to explore because
of strong phonon contributions to the absorption, which mask the
evolution of electronic resonances. Nevertheless, a slight deviation
of the experimental data from a model of an ideal, gapless and
intrinsic MCT becomes apparent.

Firstly, we concentrate on two, most pronounced interband
transitions in the region around $25$~meV and conclude that they
appear at higher energies than those expected from simple
calculations using Eq.~(\ref{LLs=}). This discrepancy points towards
an MCT with a small but still non-zero gap $E_g$. Setting
$E_g=4\;\mbox{meV}$ in calculations we satisfactorily improve the
data modelling (see solid lines in Fig.~\ref{Fig5}).

The second relevant observation is that the interband transitions,
which are strong at higher energies, rather suddenly weaken in the
limit of low magnetic fields. At the same time the lowest energy
interband transition transforms into a resonance which follows a
linear rather than $\sqrt{B}$ dependence in small
magnetic fields. These effects indicate that our structure is not
perfectly intrinsic but characterized by a nonzero electron
concentration (Fermi energy $E_F$ in the conduction band). Indeed,
if the electron concentration is not zero, each interband Landau
level transition must be (Pauli) blocked at sufficiently low
fields, when the corresponding electronic Landau level crosses
below the Fermi energy. Our interband transitions are barely seen
at energies below $15$~meV, see Fig.~\ref{Fig5}. This allows us to estimate
$E_F\approx 15-17$~meV, and in consequence also the electron
concentration  $n=\frac{1}{3\pi^2 v
^3\hbar^3}[E_F(E_F-E_g)]^{3/2}=(2-3)\times10^{14}$~cm$^{-3}$
(for $E_g=4$~meV estimated above).

The presence of free electrons in our structure explains also its spectral
response at low energies (below $15$~meV). This response is due to
classical cyclotron resonance absorption at low
magnetic fields, which transforms with increasing $B$ into intraband transitions between
adjacent Landau levels and ends up as a transitions from the flat band
into L$_{1,1,\uparrow}$ level, see Fig.~\ref{Fig3}(a), when $E_F$ is locked
at the bottom of L$_{1,1,\uparrow}$ level. It is worth noticing that, independently of
parabolic or linear electronic dispersions, the classical
cyclotron resonance is linear with the magnetic field, invoking a
$m_c=E_F/v^2$ effective mass in case of linear dispersion
relations~\cite{WitowskiPRB10,CrasseeNaturPhys11,OrlitaNJP12}. In
our case, a small gap is present and the cyclotron mass
becomes $m_c=(E_F-E_g/2)/v^2 \approx 2\times 10^{-3}\,m_0$. This value corresponds well
to the cyclotron mass, $m_c=(1.5\pm0.5)\times10^{-3}\,m_0$ derived directly
from the slope of cyclotron resonance absorption in the limit of
low magnetic fields.


\vspace{1mm}

In conclusion, we have observed a new type of 3D massless fermions, which extend the currently known family of 3D massless particles. These ``Kane fermions''
share many features with previously discussed Weyl and Dirac fermions: optical absorption which is linear in frequency, Landau levels and their Zeeman splitting, which are
proportional to $\sqrt{B}$ and rigidly related to each other. However, the relation between the spin and orbital splitting for massless Kane fermions is different from that for the ultra-relativistic Dirac electrons or Weyl fermions, which is one manifestation of their inequivalence. An important difference from massless electrons in Weyl semimetals is that the massless fermions in MCT are not protected by symmetry or topology; rather, we have engineered the conical dispersion by fine tuning a system parameter, cadmium concentration, which is
extremely homogeneous over a macroscopic thickness. This lack of protection, in fact, may represent a major advantage for potential applications: the very robustness of semimetals protected by symmetry or topology makes them hard to manipulate (e.g., to introduce a small controllable gap), while the band structure of the MCT can be engineered at will, as our work shows.
The high degree of technological control over this material opens further perspectives for its use in electronic devices, where one could benefit from the peculiar
properties of massless fermions, such as the suppressed backscattering (Klein paradox) and the related inefficient Auger-type recombination.

While our paper was under review, we have learned about several recent preprints,\cite{BorisenkoCM13,NeupaneCM13,LiuCM13}
where related the issues are discussed.

\vspace{0.5cm}

{\small
\noindent\textbf{Methods}
\vspace{1mm}

The sample was grown using standard molecular-beam epitaxy on a
(013)-oriented semi-insulating GaAs substrate. The growth sequence
started with ZnTe and CdTe transition regions, followed by the MCT
epilayer with gradually changing cadmium content~$x$ (see
Supplementary Information). The prepared MCT layer contains a
region with $x\approx{0.17}$ of thickness
$d\approx{3}.2\:\mu\mbox{m}$.

{\small The absorption coefficient of MCT was measured in the transmission
configuration. A macroscopic area of the sample of about
$4\:\mbox{mm}^2$ was exposed to the radiation of a globar or
mercury lamp, which was analyzed by a Fourier transform
spectrometer, and via light-pipe optics delivered to the sample
placed either in a superconducting solenoid or resistive coil. At
low fields, the correction for the remanent field of the solenoid
has been made. The transmitted light was detected by a composite
bolometer placed directly below the sample, kept at a temperature
of 1.8~K. The sample transmission $T_B$ at a given magnetic field $B$,
was normalized by the substrate transmission,
$T_S$, measured in the absence of MCT. The absorption
coefficient $\lambda_B$ was
determined from the relation
$T_B/T_S=\exp(-\lambda_B{d})$. This relation neglects the
dielectric mismatch between MCT and GaAs. A significant mismatch
would result in additional reflection and would produce a constant
vertical shift of the curve in Fig.~\ref{Fig2}, so the straight
line would not pass through the origin. The fact that it does,
shows that the dielectric mismatch is indeed negligible. This
implies $\vep_\infty\approx{6}$, when the realistic cut-off energy
$\Omega=1.5\,\mbox{eV}$ is assumed~\cite{HassPRB83}.

\vspace{2mm}
\noindent\textbf{Acknowledgements}
\vspace{1mm}

The authors acknowledge helpful discussions with T. Brauner,
R. Grill, M. Grynberg, A.~A.~Nersesyan, V. Nov\'{a}k, M. L. Sadowski,
and W. Zawadzki. The work has been supported by ERC project MOMB and by EuroMagNET II under the EU Contract
No. 228043.

\vspace{2mm}
\noindent\textbf{Author contributions}
\vspace{1mm}

The experiment was proposed by M.O. and M.P., underlying theory was formulated by D.M.B.
The sample growth was performed by N.N.M. and S.A.D. The sample was characterized by M.Z., F.T., W.K. and V.I.G.
Magneto-optical experiments were performed by M.O., G.M., M.Z,  P.N., C.F. and A.L.B.
All coauthors discussed the data. M.O., M.P. and D.M.B. wrote the manuscript.}


\begin{thebibliography}{10}

\bibitem{CharlierRMP07}
Charlier, J.-C., Blase, X., and Roche, S.
Electronic and transport properties of nanotubes.
\newblock {\em Rev. Mod. Phys.}{ \bf 79}, 677--732 (2007).

\bibitem{NovoselovNature05}
Novoselov, K.~S. \emph{et al.}
Two-dimensional gas of massless Dirac fermions in graphene.
\newblock {\em Nature}{ \bf 438}, 197 (2005).

\bibitem{ZhangNature05}
Zhang, Y.~B., Tan, Y.~W., Stormer, H.~L., and Kim, P.
Experimental observation of the quantum Hall effect and Berrys phase in graphene.
\newblock {\em Nature}{ \bf 438}, 201 (2005).

\bibitem{KonigScience07}
K\"{o}nig, M. \emph{et al.}
Quantum spin Hall insulator state in HgTe quantum wells.
\newblock {\em Science}{ \bf 318}(5851), 766--770 (2007).

\bibitem{HasanRMP10}
Hasan, M.~Z. and Kane, C.~L.
Colloquium: Topological insulators.
\newblock {\em Rev. Mod. Phys.}{ \bf 82}, 3045--3067 (2010).

\bibitem{YoungPRL12}
Young, S.~M. \emph{et al.}
Dirac semimetal in three dimensions.
\newblock {\em Phys. Rev. Lett.}{ \bf 108}, 140405 (2012).

\bibitem{WanPRB11}
Wan, X., Turner, A. M., Viswanath, A., and Savrasov, S. Y.
Topological semimetal and Fermi-arc surface states in the
electronic structure of pyrochlore iridates.
\textit{Phys. Rev. B} \textbf{83}, 205101 (2011).

\bibitem{YangPRB11}
Yang, K.-Y., Lu, Y.-M., and Ran, Y.
Quantum Hall effects in a Weyl semimetal:
Possible application in pyrochlore iridates.
\textit{Phys. Rev. B} \textbf{84}, 075129 (2011).

\bibitem{WangPRB12}
Wang, Z. \emph{et al.}
Dirac semimetal and topological phase transitions in ${A}_{3}$Bi ($A=\text{Na}$, K, Rb).
\newblock {\em Phys. Rev. B}{ \bf 85}, 195320 (2012).


\bibitem{SteinbergCM13}
Steinberg, J. A. \emph{et al.}
Bulk Dirac points in distorted spinels.
arXiv:1309.5967 (2013).

\bibitem{SinghPRB12}
Singh, B. \emph{et al.}
Topological electronic structure and Weyl semimetal in the TlBiSe${}_{2}$ class of semiconductors.
\newblock {\em Phys. Rev. B}{ \bf 86}, 115208 (2012).


\bibitem{XuScience11}
Xu, S.-Y. \textit{et al.}
Topological Phase Transition and Texture Inversion
in a Tunable Topological Insulator.
\textit{Science} \textbf{332}, 560--564 (2011).

\bibitem{SatoNaturePhys11}
Sato, T. \textit{et al.}
Unexpected mass acquisition of Dirac fermions
at the quantum phase transition of a topological insulator.
\textit{Nature Phys.} \textbf{7}, 840--844 (2011).

\bibitem{ZawadzkiAinP74}
Zawadzki, W.
Electron transport phenomena in small-gap semiconductors.
\newblock {\em Advances in Physics}{ \bf 23}, 435--522 (1974).

\bibitem{HarmanPRL61}
Harman, T. C. \emph{et al.}
Low Electron Effective Masses and Energy Gap in Cd$_x$Hg$_{1-x}$Te.
\newblock {\em Phys. Rev. Lett.}{ \bf 7}, 403 (1961).

\bibitem{McCombeSSC70}
McCombe, B. D., Wagner, R., and Prinz, G.
Infrared pulsed gas laser studies of combined resonance and cyclotron-phonon resonance in Hg$_{1-x}$Cd$_x$Te.
\newblock {\em Solid State Communications}{ \bf 8}, 1687 -- 1691 (1970).

\bibitem{GrovesSSC71}
Groves, S. H., Harman, T. C., and Pidgeon, C. R.
Interband magnetoreflection of Hg$_{1-x}$Cd$_x$Te.
\newblock {\em Solid State Communications}{ \bf 9},  451 –- 455 (1971).

\bibitem{GuldnerPPSA77}
Guldner, Y., Rigaux, C., Mycielski, A., and Couder, Y.
Magnetooptical investigation of Hg$_{1-x}$Cd$_x$Te mixed crystals II. Semiconducting configuration and semimetal $\rightarrow$ semiconductor transition.
\newblock {\em physica status solidi (b)}{ \bf 82}, 149--158 (1977).

\bibitem{WeilerSaS81}
Weiler, M.~H.
\newblock In {\em Defects, (HgCd)Se, (HgCd)Te, }Willardson, R.~K. and Beer,
  A.~C., editors, volume~16 of {\em Semiconductors and Semimetals},  119 --
  191. Elsevier (1981).

\bibitem{BernevigScience06}
Bernevig, B.~A., Hughes, T.~L., and Zhang, S.-C.
Quantum spin Hall effect and topological phase transition in HgTe quantum wells.
\newblock {\em Science}{ \bf 314}, 1757--1761 (2006).

\bibitem{KaneJPCS57}
Kane, E.~O.
Band structure of indium antimonide.
\newblock {\em Journal of Physics and Chemistry of Solids}{ \bf 1}, 249 -- 261 (1957).

\bibitem{NovikPRB05}
Novik, E.~G. \emph{et al.}
Band structure of semimagnetic ${\mathrm{Hg}}_{1-y}{\mathrm{Mn}}_{y}\mathrm{Te}$ quantum wells.
\newblock {\em Phys. Rev. B}{ \bf 72}, 035321 (2005).

\bibitem{Cardona}
Yu, P.~Y. and Cardona, M.
Fundamentals of Semiconductors.
\newblock (Springer, Heidelberg, 1996).

\bibitem{Balents}
Eq. 4 is valid for the intrinsic material at zero
temperature, $T=0$. It has been noted recently [see Burkov, A. A. and Balents, L., Weyl Semimetal in a Topological Insulator Multilayer.
\emph{Phys. Rev. Lett.} \textbf{107}, 127205 (2012)] that in the intrinsic
case (zero Fermi energy) the limits $\omega\to{0}$ and $T\to{0}$
do not commute. As discussed below, our sample is not fully
intrinsic (some small residual population in the conduction band
is present), so the limit $\omega\to{0}$ is regularized by this
residual population.


\bibitem{Katsnelson2006}
Katsnelson, M.~I., Novoselov, K.~S., and Geim, A.~K.
Chiral tunnelling and the Klein paradox in graphene.
\newblock {\em Nature Phys.}{ \bf 2}, 620--625 (2006).

\bibitem{YoungNaturePhys09}
Young, A.~F. and Kim, P.
Quantum interference and Klein tunnelling in graphene heterojunctions.
\newblock {\em Nature Phys.}{ \bf 5}, 222--226 (2009).

\bibitem{KuzmenkoPRL08}
Kuzmenko, A.~B. \emph{et al.}
Universal optical conductance of graphite.
\newblock {\em Phys. Rev. Lett.}{ \bf 100}, 117401 (2008).

\bibitem{NairScience08}
Nair, R.~R. \emph{et al.}
Fine structure constant defines visual transparency of graphene.
\newblock {\em Science}{ \bf 320}, 1308 (2008).

\bibitem{YuCardona}
Yu, P.~Y. and Cardona, M.
\newblock {\em Fundamentals of Semiconductors}.
\newblock Springer, Heidelberg (1999).

\bibitem{LL4}
Berestetskii, V.~B., Lifshitz, E.~M., and Pitaevskii, L.~P.
\newblock (Pergamon, Oxford, 1971).

\bibitem{AshbyPRB13}
Ashby, P.~E.~C. and Carbotte, J. P.
Magneto-optical conductivity of Weyl semimetals.
\newblock {\em Phys. Rev. B}{ \bf 87}, 245131 (2013).

\bibitem{ZhuPRB11}
Zhu, Z. \emph{et al.}
Angle-resolved Landau spectrum of electrons and holes in bismuth.
\newblock {\em Phys. Rev. B}{ \bf 84}, 115137 (2011).

\bibitem{OrlitaPRL08}
Orlita, M. \emph{et al.}
Dirac Fermions at the $H$ Point of Graphite: Magnetotransmission Studies
\newblock {\em Phys. Rev. Lett.}{ \bf 100}, 136403 (2008).

\bibitem{WitowskiPRB10}
Witowski, A.~M. \emph{et al.}
Quasiclassical cyclotron resonance of Dirac fermions in highly doped graphene.
\newblock {\em Phys. Rev. B}{ \bf 82}, 165305 (2010).

\bibitem{CrasseeNaturPhys11}
Crassee, I. \emph{et al.}
Giant Faraday rotation in single- and multilayer graphene.
\newblock {\em Nature Phys.}{ \bf 7}, 48--51 (2011).

\bibitem{OrlitaNJP12}
Orlita, M. \emph{et al.}
Classical to quantum crossover of the cyclotron resonance in graphene: a study of the strength of intraband absorption.
\newblock {\em New J. of Phys.}{ \bf 14}, 095008 (2012).

\bibitem{HassPRB83}
Hass, K.~C., Ehrenreich, H., and Velick\'y, B.
Electronic structure of ${\mathrm{Hg}}_{1-x}{\mathrm{Cd}}_{x}\mathrm{Te}$.
\newblock {\em Phys. Rev. B}{ \bf 27}, 1088--1100 (1983).

\bibitem{NeupaneCM13}
Neupane, M. \textit{et al.}
Observation of a topological 3D Dirac semimetal phase in high-mobility Cd$_3$As$_2$.
arXiv:1309.7892 (2013).

\bibitem{BorisenkoCM13}
Borisenko, S. \textit{et al.}
Experimental Realization of a Three-Dimensional Dirac Semimetal.
arXiv:1309.7978 (2013).

\bibitem{LiuCM13}
Liu, Z. K. \textit{et al.}
Discovery of a Three-dimensional Topological Dirac Semimetal, Na$_3$Bi.
arXiv:1310.0391 (2013).
\end{thebibliography}

\newpage

\begin{widetext}
\begin{center}
{\large
\textbf{Supplementary Information for}\\\vspace{2mm}
\emph{3D massless Kane fermions observed in a zinc-blende crystal}\\\vspace{2mm}
by M. Orlita, D. M. Basko, M. S. Zholudev, F. Teppe, W. Knap, V. I. Gavrilenko, N. N. Mikhailov,
S. A. Dvoretskii, P. Neugebauer, C. Faugeras, A.-L. Barra, G. Martinez, and M. Potemski}
\end{center}

\section{Sample structure}

The studied sample was grown using the standard MBE technique
on the (013)-oriented semi-insulating GaAs substrate. The growth
sequence started with ZnTe and CdTe transition (buffer) regions, followed
by the MCT epilayer with gradually changing cadmium content~$x$. The
profile of cadmium content is shown in Fig.~\ref{Supp1}. The prepared MCT layer contains
a region with $x\approx{0.17}$ of thickness $d\approx{3}.2\:\mu\mbox{m}$.
The cadmium profile has been controlled during growth using in situ single wavelength ellipsometry, see, e.g., Ref.~S1.

\begin{figure}[h!]
      \includegraphics[width=8cm]{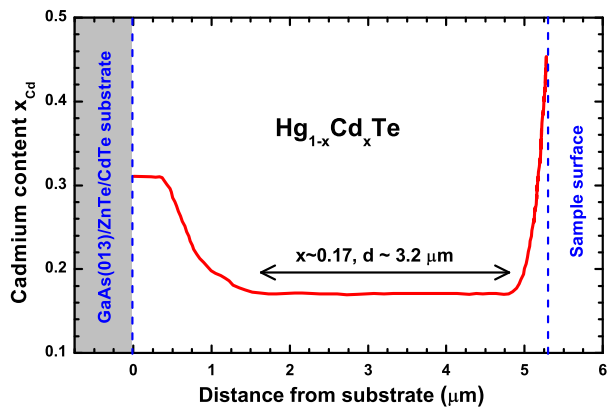}
      \caption{\label{Supp1}
The profile of cadmium content in the studied sample.}
\end{figure}

\section{Kane model and the effective band Hamiltonian}

In zinc-blende semiconductors the orbital degeneracies of the
conduction and valence bands are 1 and 3, respectively. At
$\vec{k}=0$ we can choose a real Bloch function $u_c(\vec{r})$ for
the conduction band, and three real functions
$u_X(\vec{r}),\,u_Y(\vec{r}),\,u_Z(\vec{r})$ for the valence band.
The function $u_c(\vec{r})$~transforms according to the identical
representation $\Gamma_1$ of the crystal group~$T_d$, while
$u_X(\vec{r}),\,u_Y(\vec{r}),\,u_Z(\vec{r})$ transform according
to the vector $\Gamma_{15}$ representation, equivalently to the
functions $x,y,z$.
Out of the three real functions $u_\alpha(\vec{r})$ one can make
linear combinations $u_m(\vec{r})$ corresponding to eigenfunctions
of the $z$-projection of the orbital angular momentum $l=1$:
\begin{equation}\begin{split}
&u_{+1}=-\frac{i}{\sqrt{2}}\,(u_X+iu_Y),\\
&u_0=iu_Z,\\
&u_{-1}=\frac{i}{\sqrt{2}}\,(u_X-iu_Y).
\end{split}\end{equation}
The spin structure of the wave functions can be accounted for by
introducing two spinors $\chi_\uparrow,\chi_\downarrow$,
corresponding to the two values of the spin projection on the
$z$~axis.
Spin-orbit interaction splits the $2(2l+1)$-fold degenerate
valence band into two subspaces corresponding to the total angular
momentum $J=1/2$ and $J=3/2$, the latter manifold corresponding to
the topmost valence band. Explicitly,
\begin{equation}\begin{split}
&\begin{array}{l}
u_{3/2;+3/2}=u_{+1}\chi_\uparrow,\\
u_{3/2;+1/2}=\sqrt{2/3}\,u_0\chi_\uparrow
+\sqrt{1/3}\,u_{+1}\chi_\downarrow,\\
u_{3/2;-1/2}=\sqrt{2/3}\,u_0\chi_\downarrow
+\sqrt{1/3}\,u_{-1}\chi_\uparrow,\\
u_{3/2;-3/2}=u_{-1}\chi_\downarrow,
\end{array}\\
&\begin{array}{l}
u_{1/2;+1/2}=\sqrt{1/3}\,u_0\chi_\uparrow
-\sqrt{2/3}\,u_{+1}\chi_\downarrow,\\
u_{1/2;-1/2}=-\sqrt{1/3}\,u_0\chi_\downarrow
+\sqrt{2/3}\,u_{-1}\chi_\uparrow,
\end{array}
\end{split}\end{equation}
It is convenient to arrange the basis vectors as
\begin{eqnarray}
&&\left(\begin{array}{cccccccc}
u_c\chi_\uparrow & u_{3/2,+3/2} & u_{3/2,-1/2} & u_{1/2,+1/2} &
u_c\chi_\downarrow & u_{3/2,-3/2} & u_{3/2,+1/2} & u_{1/2,-1/2}
\end{array}\right)\\
&&=\left(\begin{array}{cccccccc}
u_c\chi_\uparrow & u_X\chi_\uparrow & u_Y\chi_\uparrow & u_Z\chi_\uparrow &
u_c\chi_\downarrow & u_X\chi_\downarrow & u_Y\chi_\downarrow & u_Z\chi_\downarrow
\end{array}\right)U,\\
&&U=\left(\begin{array}{cccccccc}
1 & 0 & 0 & 0 & 0 & 0 & 0 & 0 \\
0 &-\sqrt{1/2}\,i & \sqrt{1/6}\,i & 0 & 0 & 0 & 0 & \sqrt{1/3}\,i\\
0 & \sqrt{1/2} & \sqrt{1/6} & 0 & 0 & 0 & 0 & \sqrt{1/3} \\
0 & 0 & 0 & \sqrt{1/3}\,i & 0 & 0 & \sqrt{2/3}\,i & 0 \\
0 & 0 & 0 & 0 & 1 & 0 & 0 & 0 \\
0 & 0 & 0 & \sqrt{1/3}\,i & 0 & \sqrt{1/2}\,i & -\sqrt{1/6}\,i & 0\\
0 & 0 & 0 & -\sqrt{1/3} & 0 &  \sqrt{1/2} & \sqrt{1/6} & 0\\
0 & 0 & \sqrt{2/3}\,i & 0 & 0 & 0 & 0 & -\sqrt{1/3}\,i
\end{array}\right),
\end{eqnarray}
then the time reversal matrix is just $\sigma_y$, the second Pauli
matrix acting in the $2\times{2}$ space made of $4\times{4}$ blocks.

In this basis, the electronic Hamiltonian at $\vec{k}=0$ is given
by
\begin{equation}
H(\vec{k}=0)=\left(\begin{array}{cccccccc}
E_g & 0 & 0 & 0 & 0 & 0 & 0 & 0 \\
0 & 0 & 0 & 0 & 0 & 0 & 0 & 0\\ 0 & 0 & 0 & 0 & 0 & 0 & 0 & 0\\
0 & 0 & 0 & -\Delta & 0 & 0 & 0 & 0 \\
0 & 0 & 0 & 0 & E_g & 0 & 0 & 0 \\
0 & 0 & 0 & 0 & 0 & 0 & 0 & 0\\ 0 & 0 & 0 & 0 & 0 & 0 & 0 & 0\\
0 & 0 & 0 & 0 & 0 & 0 & 0 & -\Delta
\end{array}\right),
\end{equation}
where the energy is counted from the top of the $J=3/2$ valence band.
$\Delta$~is the spin-orbit splitting between the $J=3/2$ and the
$J=1/2$ valence bands. The band gap is parametrized by
$E_g\approx(x-x_c)\cdot{1}.9\:\mbox{eV}$, see Ref. S2.
Since $E_g<0$ at $x<x_c$, the semimetallic MCT is sometimes called
a negative-gap semiconductor.

The linear in $\vec{k}$ terms in the effective band Hamiltonian are
obtained in the first order of the $\vec{k}\cdot\vec{p}$ perturbation
theory. The momentum matrix elements between the conduction and the
valence band Bloch functions are determined by
\begin{equation}
\int{u}_\alpha(\vec{r})\,
\frac{\partial{u}_c(\vec{r})}{\partial{x}_\beta}\,
d^3\vec{r}=P\delta_{\alpha\beta},\quad
\end{equation}
where $P$~is the Kane's matrix element, and $2P^2/m_0\equiv{E}_P$
is called Kane's energy ($m_0$~is the free electron mass).

The effective Hamiltonian to $O(k)$ is given by:
\begin{equation}\begin{split}\label{H8x8=}
H(\vec{k})={}&{}H(\vec{k}=0)+U^\dagger\left[
\frac{P}{m_0}\left(\begin{array}{cccc}
0 & ik_x & ik_y & ik_z \\ -ik_x & 0 & 0 & 0 \\
-ik_y & 0 & 0 & 0 \\ -ik_z & 0 & 0 & 0
\end{array}\right)\otimes
\left(\begin{array}{cc} 1 & 0 \\
0 & 1 \end{array}\right)\right]U=\\
={}&{}\left(\begin{array}{cccccccc}
E_g & vk_+\sqrt{3}/2 & -vk_-/2 & -vk_z/\sqrt{2} &
0 & 0 & -vk_z & -vk_-/\sqrt{2} \\
vk_-\sqrt{3}/2 & 0 & 0 & 0 & 0 & 0 & 0 & 0 \\
-vk_+/2 & 0 & 0 & 0 & -vk_z & 0 & 0 & 0 \\
-vk_z/\sqrt{2} & 0 & 0 & -\Delta & -vk_-/\sqrt{2} & 0 & 0 & 0 \\
0 & 0 & -vk_z & -vk_+/\sqrt{2} &
E_g & -vk_-\sqrt{3}/2 & vk_+/2 & vk_z/\sqrt{2} \\
0 & 0 & 0 & 0 & -vk_+\sqrt{3}/2 & 0 & 0 & 0 \\
-vk_z & 0 & 0 & 0 & vk_-/2 & 0 & 0 & 0 \\
-vk_+/\sqrt{2} & 0 & 0 & 0 & vk_z/\sqrt{2} & 0 & 0 & -\Delta
\end{array}\right),
\end{split}\end{equation}
where $v\equiv\sqrt{3/2}\,P/m_0$, and $k_\pm\equiv{k}_x\pm{i}k_y$.
This Hamiltonian obeys the time-reversal symmetry,
$\sigma_yH^*(\vec{k})\sigma_y=H(-\vec{k})$, where $\sigma_y$ is the
second Pauli matrix acting in the $2\times{2}$ space made of
$4\times{4}$ blocks. The eigenvalues of the Hamiltonian~(\ref{H8x8=})
can be found from the equation
\begin{equation}
\det(H-\ep)=
\ep^2\left\{\ep^3+(\Delta-E_g)\ep^2-
[E_g\Delta+(3/2)v^2k^2]\ep-v^2k^2\Delta\right\}^2=0.
\end{equation}
They do not depend on the direction of~$\vec{k}$.

In the limit of large~$\Delta$, the Hamiltonian can be easily
projected on the subspace, orthogonal to the the split-off band.
If we are not interested in terms quadratic in~$\vec{k}$, the
projection is done by simply eliminating the fourth and the eight
row and column of the matrix in Eq.~(\ref{H8x8=}):
\begin{equation}\begin{split}\label{H6x6=}
H(\vec{k})=\left(\begin{array}{cccccc}
E_g & vk_+\sqrt{3}/2 & -vk_-/2  & 0 & 0 & -vk_z \\
vk_-\sqrt{3}/2 & 0 &  0 & 0 & 0 & 0 \\
-vk_+/2 & 0 & 0 & -vk_z & 0 & 0 \\
0 & 0 & -vk_z & E_g & -vk_-\sqrt{3}/2 & vk_+/2 \\
0 & 0 & 0 & -vk_+\sqrt{3}/2 & 0 & 0 \\
-vk_z & 0 & 0 & vk_-/2 & 0 & 0 \end{array}\right).
\end{split}\end{equation}
This matrix has three doubly-degenerate eigenvalues:
\begin{equation}
\ep_\vec{k}=0,\quad
\ep_\vec{k}=\frac{E_g}{2}\pm\sqrt{\frac{E_g^2}4+v^2k^2}.
\end{equation}
The eigenvalue $\ep=0$ corresponds to the heavy-hole band,
which in this approximation is completely flat.

Let us see how the existence of the flat band follows from the
property $U_cH(\vec{k})U_c=-H(\vec{k})$, with $H(\vec{k})$ given
by Eq.~(\ref{H6x6=}) and $U_c=\mathrm{diag}(1,-1,-1,1,-1,-1)$.
Consider the general situation: an $(n+m)\times(n+m)$
matrix~$A$, anticommuting with a matrix $U_c$ which has
$m$~eigenvalues equal to~1, and $n$~eigenvalues equal to $-1$,
and $m<n$. Let us work in the basis of the eigenvectors of~$U_c$,
which are arranged in such an order that
$U_c=\mathrm{diag}(-1,\ldots,-1,1,\ldots,1)$. The condition
$U_cAU_c=-A$ implies that in this basis the matrix~$A$ has the
following block structure:
\begin{equation}
A=\left(\begin{array}{cc} 0_{n\times{n}} & A_{n\times{m}}' \\
A_{m\times{n}}'' & 0_{m\times{m}}
\end{array}\right).
\end{equation}
Consider now the $n$-dimensional subspace of column vectors
$\underline{x}=(x_1,x_2,\ldots,x_n,0,\ldots,0)^T$. All these
vectors satisfy the first $n$~equations of the linear system
$A\underline{x}=0$. The remaining $m$ equations leave an
$(n-m)$ dimensional subspace of solutions $A\underline{x}=0$,
which corresponds to the zero eigenvalue of~$A$ with
multiplicity~$n-m$.

\section{Optical absorption at zero magnetic field}

Let us start from the standard expression for the optical
conductivity, obtained from the Kubo formula for the response
of the current to the monochromatically oscillating vector
potential:
\begin{equation}\label{sigmaij=}
\sigma_{ij}(\omega)=-ie^2\int\frac{d^3\vec{k}}{(2\pi)^3}
\sum_{l,l'=1}^6\frac{f_{l,\vec{k}}-f_{l',\vec{k}}}%
{\ep_{l,\vec{k}}-\ep_{l',\vec{k}}}\,
\frac{\langle{l},\vec{k}|v_i|l',\vec{k}\rangle
\langle{l}',\vec{k}|v_j|l,\vec{k}\rangle}%
{\omega-\ep_{l',\vec{k}}+\ep_{l,\vec{k}}+i0^+}.
\end{equation}
Here $l,l'=1,\ldots,6$ label the eigenstates of $H(\vec{k})$
which is given by Eq.~(\ref{H6x6=}), $f_{l,\vec{k}}$ are the
occupations of these eigenstates, and the velocity matrices
are $v_i=\partial{H}(\vec{k})/\partial{k}_i=vJ_i$, where
$i,j=x,y,z$ label the Cartesian components.

To calculate the velocity matrix elements, we note that the
projection of the vector $\vec{J}$ on an arbitrary direction
$\vec{n}=(\sin\vartheta\cos\varphi,\sin\vartheta\sin\varphi,
\cos\vartheta)$, determined by the spherical angles
$\vartheta,\varphi$, can be related to~$J_z$ by a rotation
\begin{eqnarray}
&&\vec{J}\cdot\vec{n}
=J_x\sin\vartheta\cos\varphi+J_y\sin\vartheta\sin\varphi
+J_z\cos\vartheta
=U_\varphi^\dagger(J_x\sin\vartheta+J_z\cos\vartheta)U_\varphi
=U_\varphi^\dagger{U}_\vartheta^\dagger{J}_z
U_\vartheta{U}_\varphi,\label{rotate=}\\
&&U_\varphi=\mathrm{diag}\left(e^{i\varphi/2},e^{3i\varphi/2},
e^{-i\varphi/2},e^{-i\varphi/2},e^{-3i\varphi/2},e^{i\varphi/2}\right),\\
&&U_\vartheta=\left(\begin{array}{cccccc}
c & 0 & 0 & s & 0 & 0 \\
0 & c^3 & \sqrt{3}cs^2 & 0 & s^3 & \sqrt{3}c^2s \\
0 & \sqrt{3}cs^2 & c^3-2cs^2 & 0 & \sqrt{3}c^2s & s^3-2c^2s \\
-s & 0 & 0 & c & 0 & 0 \\
0 & -s^3 & -\sqrt{3}c^2s & 0 & c^3 & \sqrt{3}cs^2 \\
0 & -\sqrt{3}c^2s & -s^3+2c^2s & 0 & \sqrt{3}cs^2 & c^3-2cs^2 \end{array}\right),\quad
c\equiv\cos\frac\vartheta{2},\quad
s\equiv\sin\frac\vartheta{2}.
\end{eqnarray}
Thus, the eigenstates $|l,\vec{k}\rangle$ for an arbitrary
direction of $\vec{k}$ can be related to those for $\vec{k}$
along~$z$ by
$|l,\vec{k}\rangle=U_\varphi^\dagger{U}_\vartheta^\dagger
|l,k,z\rangle$,
where $\vartheta,\varphi$ are the spherical angles of~$\vec{k}$.

By symmetry, the tensor structure of the conductivity is trivial,
$\sigma_{ij}(\omega)=\sigma(\omega)$. This can also be shown by
the direct calculation, whose details we do not give, but which
is fully analogous to the one given below. We calculate just one
component, $\sigma_{zz}$. Since the energies $\ep_{l,\vec{k}}$
depend only on~$|\vec{k}|$, we can integrate over the angles
using Eq.~(\ref{rotate=}):
\begin{equation}
\mathcal{J}_{ll'}=\int\sin\vartheta\,d\vartheta\,{d}\varphi
\left|\langle{l},\vec{k}|J_z|l',\vec{k}\rangle\right|^2
=\frac{8\pi}3\left|\langle{l},k,z|J_x|l',k,z\rangle\right|^2
+\frac{4\pi}3\left|\langle{l},k,z|J_z|l',k,z\rangle\right|^2.
\end{equation}
The eigenvectors of the Hamiltonian (\ref{H6x6=}) for $\vec{k}$
along the $z$~axis are (in the order of decreasing energy)
\[
\left(\begin{array}{c}
\mathcal{S} \\ 0 \\ 0 \\ 0 \\ 0 \\ -\mathcal{C}
\end{array}\right),
\left(\begin{array}{c}
0 \\ 0 \\ \mathcal{S} \\ -\mathcal{C} \\ 0 \\ 0
\end{array}\right),
\left(\begin{array}{c}
0 \\ 1 \\ 0 \\ 0 \\ 0 \\ 0
\end{array}\right),
\left(\begin{array}{c}
0 \\ 0 \\ 0 \\ 0 \\ 1 \\ 0
\end{array}\right),
\left(\begin{array}{c}
\mathcal{C} \\ 0 \\ 0 \\ 0 \\ 0 \\ \mathcal{S}
\end{array}\right),
\left(\begin{array}{cccccc}
0 \\ 0 \\ \mathcal{C} \\ \mathcal{S} \\ 0 \\ 0
\end{array}\right),
\]
where we have denoted
\[
\mathcal{C}=\cos\left(\frac{\phi_g}{2}+\frac{\pi}{4}\right),\quad
\mathcal{S}=\sin\left(\frac{\phi_g}{2}+\frac{\pi}{4}\right),\quad
\phi_g\equiv\arcsin\frac{E_g/2}{\sqrt{E_g^2/4+v^2k^2}}.
\]
This gives
\begin{equation}
\mathcal{J}_{ll'}=\frac{\pi}{3}\left(\begin{array}{cccccc}
4c_g^2   & 2s_g^2   & 3(1+s_g) & 0        & 4s_g^2   & 2c_g^2   \\
2s_g^2   & 4c_g^2   & 0        & 3(1-s_g) & 2c_g^2   & 4s_g^2   \\
3(1+s_g) & 0        & 0        & 0        & 3(1-s_g) & 0        \\
0        & 3(1-s_g) & 0        & 0        & 0        & 3(1+s_g) \\
4s_g^2   & 2c_g^2   & 3(1-s_g) & 0        & 4c_g^2   & 2s_g^2   \\
2c_g^2   & 4s_g^2   & 0        & 3(1+s_g) & 2s_g^2   & 4c_g^2
\end{array}\right),\quad c_g=\cos\phi_g, \quad s_g=\sin\phi_g.
\end{equation}
Substituting this into Eq.~(\ref{sigmaij=}), we finally obtain
\begin{eqnarray}
&&\Re\sigma(\omega>0)=\frac{\pi^2}{3}\frac{e^2}{v\omega}
\int\limits_0^\infty\frac{\xi^2\,d\xi}{8\pi^3}
\left[6\,\delta\!\left(\frac{E_g}{2}+\sqrt{\frac{E_g^2}4+\xi^2}-\omega\right)
+4\left(1+\frac{E_g^2}{E_g^2+4\xi^2}\right)
\delta\!\left(\sqrt{E_g^2+4\xi^2}-\omega\right)\right]=\nonumber\\
&&\qquad\qquad=\frac{e^2}{4\pi{v}}
\left[\theta(2\omega-|E_g|-E_g)
\left(1-\frac{E_g}{2\omega}\right)\sqrt{\omega^2-\omega{E}_g}
+\frac{1}{16}\,\theta(\omega-|E_g|)
\left(1+\frac{E_g^2}{\omega^2}\right)
\sqrt{\omega^2-E_g^2}\right].
\end{eqnarray}
For the gapless case, $E_g=-$, we obtain $\Re\sigma(\omega>0)=\frac{13}{12}\omega\frac{e^2}{v\pi}$.
The imaginary part of the dielectric function, $\varepsilon(\omega)=1+{i}\sigma(\omega)/(\varepsilon_0\omega)$,
then becomes $\Im \varepsilon (\omega >0) = \frac{13}{12}\frac{c}{v}\alpha$, where $\alpha$ is the fine structure constant.

\section{Landau levels}

In the presence of a magnetic field, described by the vector
potential in the Landau gauge $A_x=-By$, $A_y=A_z=0$, we make
the standard Peierls substitution $\vec{p}\to\vec{p}-e\vec{A}$
in the Hamiltonian~(\ref{H6x6=}), and seek the eigenstates in
the form
\begin{equation}\label{psixy=}
\psi(x,y)=e^{ip_xx}\left(\begin{array}{cccccc}
x_\uparrow\Phi_{n-1} & y_\uparrow\Phi_n & z_\uparrow\Phi_{n-2} &
x_\downarrow\Phi_{n-2} & y_\downarrow\Phi_{n-3} &
z_\downarrow\Phi_{n-1}
\end{array}\right)^T,
\end{equation}
where $\Phi_n=\Phi_n(y+p_xl_B^2)$ are the harmonic oscillator
wave functions and $l_B$ is the magnetic length. It can be checked
directly that the form~(\ref{psixy=}) is preserved upon action
on $\psi(x,y)$ by the Hamiltonian.
The coefficients satisfy the following linear system
(we denote $\zeta\equiv{p}_zl_B$ for brevity):
\begin{equation}\label{LLsys=}
\begin{array}{r}
\displaystyle
\frac{E_g-\ep}{v/l_B}\,x_\uparrow+\sqrt{\frac{3n}2}\,y_\uparrow
-\sqrt{\frac{n-1}2}\,z_\uparrow-\zeta{z}_\downarrow=0,\\
\displaystyle
\sqrt{\frac{3n}2}\,x_\uparrow-\frac{\ep}{v/l_B}\,y_\uparrow=0,\\
\displaystyle
-\sqrt{\frac{n-1}2}\,x_\uparrow-\frac{\ep}{v/l_B}\,z_\uparrow
-\zeta{x}_\downarrow=0,\\
\displaystyle
-\zeta{z}_\uparrow+\frac{E_g-\ep}{v/l_B}\,x_\downarrow
-\sqrt{\frac{3(n-2)}2}\,y_\downarrow+\sqrt{\frac{n-1}2}\,z_\downarrow=0,\\
\displaystyle
-\sqrt{\frac{3(n-2)}2}\,x_\downarrow-\frac\ep{v/l_B}\,y_\downarrow=0,\\
\displaystyle
-\zeta{x}_\uparrow+\sqrt{\frac{n-1}2}\,x_\downarrow
-\frac\ep{v/l_B}\,z_\downarrow=0.
\end{array}
\end{equation}
Its analysis is especially simple at $p_z=0$, when the system is
split into two decoupled $3\times{3}$ blocks for
$x_\uparrow,y_\uparrow,z_\uparrow$ and
$x_\downarrow,y_\downarrow,z_\downarrow$, respectively. It is
convenient to shift $n-1\to{n}$ in the ``$\downarrow$'' block.
In each block, the Landau levels
can be labeled by $n=0,1,2,\ldots$, $\zeta=-1,0,1$. At $n=1$,
only $\zeta=\pm{1}$ are allowed, while at $n=0$ only $\zeta=0$ exists:
\begin{eqnarray}
&&\ep_{n,\zeta,\uparrow\downarrow}=\zeta^2\,\frac{E_g}{2}
+\zeta\sqrt{\frac{E_g^2}4+\frac{v^2}{2l_B^2}\,(4n-2\pm{1})},\\
&&\psi_{n>1,0,\uparrow}=\frac{1}{\sqrt{4n-1}}
\left(\begin{array}{c} 0 \\
\sqrt{n-1}\,\Phi_n \\ \sqrt{3n}\,\Phi_{n-2}\end{array}\right),\quad
\psi_{0,0,\uparrow}=\left(\begin{array}{c} 0 \\
\phi_0 \\ 0 \end{array}\right),\\
&&\psi_{n>0,\pm{1},\uparrow}=\frac{1}{\sqrt{\ep^2+(2n-1/2)(v/l_B)^2}}
\left(\begin{array}{c} -\ep\,\Phi_{n-1} \\
-\sqrt{3n/2}\,(v/l_B)\,\Phi_n \\
\sqrt{(n-1)/2}\,(v/l_B)\,\Phi_{n-2}\end{array}\right),\\
&&\psi_{n>1,0,\downarrow}=\frac{1}{\sqrt{4n-3}}
\left(\begin{array}{c} 0 \\
\sqrt{n}\,\Phi_{n-2} \\ \sqrt{3(n-1)}\,\Phi_n\end{array}\right),\quad
\psi_{0,0,\downarrow}=\left(\begin{array}{c} 0 \\ 0 \\
\Phi_0 \end{array}\right),\\
&&\psi_{n>0,\pm{1},\downarrow}=\frac{1}{\sqrt{\ep^2+(2n-3/2)(v/l_B)^2}}
\left(\begin{array}{c} -\ep\,\Phi_{n-1} \\
\sqrt{3(n-1)/2}\,(v/l_B)\,\Phi_{n-2} \\
-\sqrt{n/2}\,(v/l_B)\,\Phi_{n}\end{array}\right).
\end{eqnarray}
The selection rules for the optical absorption at $p_z=0$ are
obtained by calculating the matrix elements of
$J_\pm=J_x\pm{i}J_y$:
\begin{equation}
\langle{n}'\zeta'\sigma'|J_+|n\zeta\sigma\rangle\propto\delta_{\sigma\sigma'}
\delta_{n',n-1}\left(1-\delta_{\zeta0}\delta_{\zeta'0}\right).
\end{equation}
At $p_z\neq{0}$, the Landau levels can be found directly from the
system~(\ref{LLsys=}):
\begin{equation}
\ep_{n,\zeta,\uparrow\downarrow}=\zeta^2\,\frac{E_g}{2}
+\zeta\sqrt{\frac{E_g^2}4+\frac{v^2}{2l_B^2}\,(4n-2\pm{1})+v^2p_z^2}.
\end{equation}
For $E_g=0$ this expression reduces to Eq.~(5) of the main text.

\vspace{1cm}

\textbf{References:}\\

[S1] N. N. Mikhailov, R. N. Smirnov, S. A. Dvoretsky, Yu. G. Sidorov, V. A. Shvets,
E. V. Spesivtsev and S. V. Rykhlitski, \textit{Int. J. Nanotechnology}
\textbf{3}, 120 (2006).\\

[S2] M. H. Weiler, in \textit{Semiconductors and Semimetals, vol. 16}
ed. by R. K. Willardson and A. C. Beer (Elsevier, 1981).

\end{widetext}

\end{document}